\documentclass[aps,prl,reprint,superscriptaddress,floatfix]{revtex4-1}
\usepackage{amsmath,amssymb,graphicx}

\begin{document}
\title{Control of the electromagnetic environment of a quantum emitter by shaping the vacuum field in a coupled-cavity system}

\author{Robert Johne}
\affiliation{Max Planck Institute for the Physics of Complex Systems, N\"othnitzer Str. 38, 01187 Dresden, Germany}

\author{Ron Schutjens}
\affiliation{COBRA Research Institute, Eindhoven University of Technology, PO Box 513, NL-5600MB Eindhoven, The Netherlands}

\author{Sartoon Fattah poor}
\affiliation{COBRA Research Institute, Eindhoven University of Technology, PO Box 513, NL-5600MB Eindhoven, The Netherlands}

\author{Chao-Yuan Jin}
\affiliation{COBRA Research Institute, Eindhoven University of Technology, PO Box 513, NL-5600MB Eindhoven, The Netherlands}

\author{Andrea Fiore}
\affiliation{COBRA Research Institute, Eindhoven University of Technology, PO Box 513, NL-5600MB Eindhoven, The Netherlands}

\date{\today}

\begin{abstract}
We propose a scheme for the ultrafast control of the emitter-field coupling rate in cavity quantum electrodynamics. This is achieved by the control of the vacuum field seen by the emitter through a modulation of the optical modes in a coupled-cavity structure. The scheme allows the on/off switching of the coupling rate without perturbing the emitter and without introducing frequency chirps on the emitted photons. It can be used to control the shape of single-photon pulses for high-fidelity quantum state transfer, to control Rabi oscillations and as a gain-modulation method in lasers. We discuss two possible experimental implementations based on photonic crystal cavities and on microwave circuits. 
\end{abstract}

\maketitle

Spontaneous emission (SE) is at the heart of quantum optics and quantum photonics. Tremendous progress has been achieved in optimizing the SE of quantum emitters (QEs) in atomic systems \cite{Raimond2001} and artificial atoms such as quantum dots \cite{Reithmaier2004,Yoshie2004,Hennessy2007} and superconducting circuits \cite{devoret2013} by placing them into resonators. These coupled cavity-QE systems have been established as nonclassical light sources \cite{shields2007} and may serve as light-matter interfaces \cite{Cirac1997,Raimond2001,Wilk2007,Ritter2012}. 

In cavities, the presence of an increased optical density of states enhances the emission and absorption properties of QEs. Typically, the QE-cavity interaction is governed by a coupling constant $g$ given by the dipole moment of the QE and the vacuum electric field associated to the cavity mode. This interaction constant is thus given by intrinsic properties of the system, which are difficult to modify. However, $g$  is a crucial parameter, since it determines the relevant timescale of the interaction. Indeed, an QE in the excited state decays into the cavity mode with a decay time $\tau_{dec} =\left( \frac{2g^2}{\kappa}\right)^{-1}$ \cite{andreani1999} in the weak coupling regime ($g\ll\kappa$), where $\kappa$ is the cavity loss rate. On the other hand, in the strong coupling regime, where the coupling constant $g$ exceeds the loss of the cavity $\kappa$, a coherent and reversible energy exchange between the cavity and the QE takes place with a characteristic timescale $\tau_r \propto 1/g$.

So far, the control of the QE-cavity interaction in the solid state has been performed mostly by tuning their spectral overlap \cite{hogele2004,faraon2007,Patel2010,midolo2012,trotta2012,thyrrestrup2013} and in the large majority of experiments on timescales much longer than the interaction time. Dynamic control is however needed for the control of the photon waveform and the establishment of QE-photon entanglement. Such dynamic control has been demonstrated recently by using a combined variation of the loss rate and the cavity field by ultrafast carrier injection \cite{jin2013}, and by ultrafast detuning in photonic crystal diodes \cite{Pagliano2014}. However, both these techniques produce a temporal variation of the cavity frequency, resulting in a frequency chirp of the emitted photons, which limits the fidelity of quantum state transfer \cite{Johne2011}.

In this letter, we propose the concept of pure vacuum field modulation as a method to control the QE-cavity interaction in real time. We demonstrate that the vacuum field of an optical mode in a given cavity can be completely suppressed by varying the frequency of two coupled lateral cavities, without producing any frequency chirp of the target mode. This enables the on/off switching of the QE-cavity interaction rate $g$, which is fundamentally different from control techniques based on controlling the QE-cavity detuning. As an example, we theoretically demonstrate the shaping of a single photon pulse into a symmetric wavepacket as a prerequisite for high-fidelity quantum state transfer \cite{Cirac1997}. Finally we discuss two experimentally feasible platforms to implement the proposed scheme. The full and direct control of the light-matter interaction constant $g$ represents powerful tool to develop advanced applications in quantum information science, e.g. switching Rabi oscillations, and it can serve as the basis for a new class of gain modulated lasers.

We first consider the coupling of three in-line cavities with a QE placed in the central cavity (called target cavity) with frequency $\omega_t$ as shown in Fig.1. Furthermore we assume that the frequency of the two outer cavities $\omega_{l,r}=\omega_{t}\pm \Delta$, named left and right control cavity in the following, can be tuned at will under the assumption that the detuning $\Delta$ is the same for both but with a different sign. The Hamiltonian of the empty three-cavity system reads ($\hbar=1$)
\begin{equation}
{H_{cc}} = \left( {\begin{array}{*{20}{c}}
{\left( {{\omega _t} + \Delta } \right) - i{\kappa _l}}&\eta &0\\
\eta &{{\omega _t} - i{\kappa _t}}&\eta \\
0&\eta &{\left( {{\omega _t} - \Delta } \right) - i{\kappa _r}}
\end{array}} \right),
\end{equation}
where $\kappa_{r,l,t}$ are the loss rate of the control cavities and target cavity, respectively, and $\eta$ denotes the coupling rate between the adjacent cavities. Neglecting the loss rates, the Hamiltonian can be exactly diagonalized, which yields three non-degenerate eigenvalues $\omega_i$ ($i=1,2,3$)
\begin{eqnarray}
\begin{array}{l}
{\omega _1} = {\omega _t}\\
{\omega _{2,3}} = {\omega _t} \pm \sqrt {2{\eta ^2} + {\Delta ^2}} .
\end{array}
\end{eqnarray}

\begin{figure}
\begin{center}
\includegraphics[width=1\linewidth]{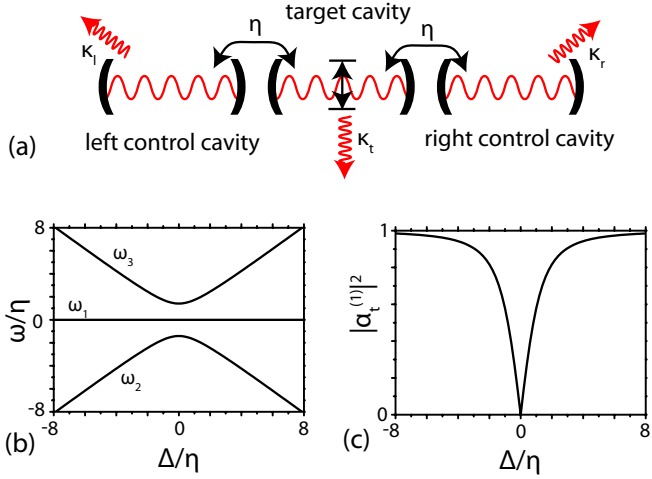}
\caption {\label{fig1} (a) Illustration of the coupled cavity scheme; (b) Calculated eigenfrequencies $\omega_i/\eta$ as a function of the dimensionless cavity detuning $\Delta/\eta$. (c) Calculated $|\alpha^{(1)}_t|^2$, which governs the modulation of the QE-cavity coupling constant $g$. The used parameters are $\kappa_{t,r,l}=0.1\eta$.}
\end{center}
\end{figure}

The most interesting eigenvalue, which we will consider in the following is $\omega_1=\omega_t$ because its frequency is independent from the detuning of the control cavities. The corresponding eigenvector reads
\begin{equation}
\vec \alpha^{(1)}  = \left( {\begin{array}{*{20}{c}}
{{\alpha _l^{(1)}}}\\
{{\alpha _t^{(1)}}}\\
{ {\alpha _r^{(1)}}}
\end{array}} \right) = \frac{1}{{\sqrt {2 + {{\left( {{\Delta  \mathord{\left/
 {\vphantom {\Delta  \eta }} \right.
 \kern-\nulldelimiterspace} \eta }} \right)}^2}} }}\left( {\begin{array}{*{20}{c}}
{ - 1}\\
{\frac{\Delta }{\eta }}\\
1
\end{array}} \right).
\end{equation}

Fig.1 (b) and (c) show the frequencies $\omega_i$ versus detuning and the target cavity fraction $|\alpha^{(1)}_t|^2$ of the mode $\omega_1$ calculated by solving the eigenvalue problem for $H_{cc}$. Interestingly, by changing the detuning of the control cavities, the target cavity fraction $\alpha^{(1)}_t$ of the eigenmode can be changed from $\alpha^{(1)}_t(\Delta=0)=0$ to $\alpha^{(1)}_t(\Delta\gg \eta)\approx 1$.  These properties provide already the main ingredient for the proposed system: The eigenmode at frequency $\omega_1=\omega_t$ can be freely engineered to have no electric field component in the target cavity or the full field intensity corresponding to a decoupled target cavity. This has drastic consequences for a QE with frequency $\omega_{e}=\omega_t$ coupled to the target cavity with strength $g<\kappa_t$. Due to its spatial position it can couple to the mode in case $\alpha^{(1)}_t>0$ or is completely decoupled if $\alpha^{(1)}_t=0$ (i.e. when the three cavities are in resonance). 

Moreover, also the effective loss rate of the modes can be changed due to the mixing of the cavity components. An approximate diagonalization ($\eta\gg\kappa_t,\kappa_c$) of Eq.(1) including the loss terms yields similar eigenvectors as given in Eq. (3). The effective loss of the coupled eigenmode $\omega_1$ can be then written as $\kappa_1=|\alpha^{(1)}_t|^2 \kappa_t+|\alpha^{(1)}_l|^2\kappa_l+|\alpha^{(1)}_r|^2\kappa_r$.
Thus, if the loss rates are equal, the tuning of the left and right control cavities allows for a pure $g$-modulation.

In order to describe the interaction of the coupled cavity system with the QE, we write the full Hamiltonian including the QE with frequency $\omega_{e}$ and the QE-target cavity coupling $g$  using the basis of the coupled modes:
\begin{equation}
{H_{eff}} = \left( {\begin{array}{*{20}{c}}
{{\omega _{e}-i\gamma}}&{{g_1}}&{{g_2}}&{{g_3}}\\
{{g_1}}&{{\omega _1} - i{\kappa _1}}&0&0\\
{{g_2}}&0&{{\omega _2} - i{\kappa _2}}&0\\
{{g_3}}&0&0&{{\omega _3} - i{\kappa _3}}
\end{array}} \right),
\end{equation} 
where the coupling rate $g_i=\alpha_t^{(i)}g$ ($i=1,2,3$) and $\alpha_t^{(i)}$ describes the target cavity fraction of each eigenmode. Assuming that $\eta \gg g$ and $\omega_{e}=\omega_t$, the QE is spectrally decoupled from the modes $\omega_{2,3}$ (as we will assume in the following).
It is immediately obvious, that by changing the detuning of the control cavities $\Delta$ one can modulate directly the effective interaction strength between the QE and the cavity mode $\alpha^{(1)}_t g$, without changing the frequency of the coupled cavity mode ($\omega_1=\omega_t$). 
The SE rate of the QE (neglecting the spontaneous decay into leaky modes given by $\gamma$) can be described in the uncoupled-cavity case by $\gamma_0=\frac{2g^2}{\kappa_t}$ \cite{andreani1999}. This obviously changes in case of the coupled cavities to $\gamma_{eff}=\frac{2|\alpha^{(1)}_t|^2g^2}{\kappa_1}$. One obtains for the ratio of the SE rates 
\begin{equation}
\frac{\gamma_{eff}}{\gamma_0}=|\alpha^{(1)}_t|^2 \frac{\kappa_t}{(|\alpha^{(1)}_t|^2\kappa_t+|\alpha^{(1)}_l|^2\kappa_l+|\alpha^{(1)}_r|^2\kappa_r)},
\end{equation}
which can be tuned from zero (all cavities in resonance and $|\alpha^{(1)}_t|=0$) to one ($\Delta\gg\eta$ and $|\alpha^{(1)}_t|\approx1$). We note that the stimulated emission rate and the modal gain in a microcavity laser are also directly related to the amplitude of the electric field at the QE's position. This vacuum field modulation technique can therefore be used to modulate the gain without directly affecting the carrier population.

The above described system can be generally used to control all spontaneous and stimulated emission processes in a microcavity. In the following, as an example, we illustrate the shaping of an emitted photon pulse from an initially inverted QE. To this aim, one has to implement a dynamic tuning of the control cavities $\Delta\rightarrow\Delta(t)$ in such a way that the typically sharp rise of the emitted single photon wavepacket can be slowed down to exactly match the time inverted decay tail. 

We simulate the dynamics using a wavefunction approach and replace the loss of the target cavity by the coupling to a quasi-continuum of modes representing a waveguide coupled to the target cavity \cite{Johne2011,Johne2012}. The Hamiltonian of the system reads
\begin{eqnarray}
H =&& \sum\limits_{i = t,r,l} {{\omega _i}a_i^ + {a_i}}  + \sum\limits_{k = 1}^N {{\omega _k}b_k^ + {b_k}}  - \sum\limits_{i = r,l}^{} {i\eta } (a_t^ + {a_i} - a_i^ + {a_t})\\ \nonumber
&& + ig(a_t^ + \sigma  - {\sigma ^ + }{a_t}) - \sqrt {\frac{\kappa_t \Delta \omega_k}{{2\pi }}} \sum\limits_{k = 1}^N {(a_t^ + {b_k} - b_k^ + {a_t})} .
\end{eqnarray}
The first term is the free evolution of the three cavity modes with operators $a_i$ and the second term describes the evolution of the quasi-continuum with mode frequencies $\omega_k$ and operators $b_k$. The remaining terms describe the cavity-cavity coupling, the QE-target cavity coupling and the target cavity-continuum coupling, respectively. 
By plugging the expansion of the wavefunction (limiting ourself to only a single excitation in the system)
\begin{eqnarray}
&&\left| \Psi  \right\rangle  = ({c_e}\left| {e,0,0,0} \right\rangle  + {c_t}\left| {g,1,0,0} \right\rangle  + {c_l}\left| {g,0,1,0} \right\rangle \\ \nonumber
 &&+ {c_r}\left| {g,0,0,1} \right\rangle ) \otimes \left| {vac} \right\rangle  + \left| {g,0,0,0} \right\rangle  \otimes \sum\limits_{k} {c_{\kappa}^{(k)}b_k^ + \left| {vac} \right\rangle } 
\end{eqnarray}
in the time dependent Schr\"odinger equation $-i\partial_t |\Psi\rangle=H_{eff}|\Psi\rangle$ using an effective Hamiltonian including the loss of the cavity modes and the decay of the QE we obtain the evolution equations for the state amplitudes in a frame rotating at $\omega_t$
\begin{eqnarray}
\label{System}
  &&{{\dot c}_{e}} = {g}{c_{t}}  - \gamma c_{e} \\ 
	&&{{\dot c}_{t}} =  - g c_{e} -\eta (c_l + c_r)+\kappa' \sum\limits_{k = 1}^N c_{\kappa}^{(k)}  \\ 
	&&{{\dot c}_{l}} = \eta c_t -i\Delta(t) c_l-\kappa_l c_l  \\ 
	&&{{\dot c}_{r}} = \eta c_t +i\Delta(t) c_r -\kappa_r c_r  \\ 
  &&{{\dot c}_{\kappa}^{(k)}} =  - i{\Delta _k}{c_{\kappa}^{(k)}} - \kappa' {c_{t}} ,
\end{eqnarray}
where $\kappa'=\sqrt {\frac{\kappa_t \Delta \omega_k}{{2\pi }}}$ is the target-cavity-continuum coupling and $\Delta \omega_k$ is the spacing of the quasi-continuum modes. The QE-dynamics are given by the amplitude $c_e$, while the target and the two control cavities are described by amplitudes $c_t$ and $c_{l,r}$, respectively. Finally the waveguide quasi-continuum is described by the amplitudes $c_{\kappa}^{(k)}$.

Starting from a inverted QE ($c_e(0)=1$) we calculate the evolution of the state amplitudes as well as the output pulse, which is given by the inverse Fourier transform of the amplitudes of the continuum $c_{\kappa}^{(k)}(T)$ at a given time $T$ much larger than the duration of the effective interaction of the QE with the cavity system \cite{Johne2011,Johne2012}. We explicitly take into account the losses of the control cavities, while the main loss channel of the target cavity is the quasi continuum coupling. Furthermore the QE is weakly coupled to the cavity mode $g<\kappa_t$.

\begin{figure}
\begin{center}
\includegraphics[width=1\linewidth]{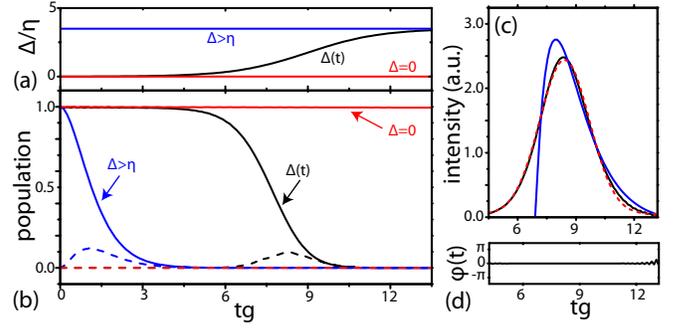}
\caption {\label{fig2} (a) Detuning $\Delta /\eta$ for three different cases: $\Delta=$const (blue), $\Delta=\Delta(t)$ (black), and $\Delta=0$ (red). (b) intracavity dynamics for the three cases. Solid lines show the QE population $|c_e|^2$ and dashed lines show the cavity photon population $|c_t|^2$. (c) Pulses emitted into the waveguide for $\Delta_1=const$ (blue) and for $\Delta_2(t)$ (black). The red dashed line is a Gaussian fit of the emitted symmetric photon pulse. (d) Phase $\phi(t)$ of the emitted photon pulse versus time for the symmetric photon pulse. The used parameters are $\left\{g,\eta, \kappa_{r,l}\right\}=\left\{0.1,10,1\right\}\kappa_t$.}
\end{center}
\end{figure}

Figure 2(a) shows the detuning of the control cavities for the three scenarios $\Delta\gg\eta$, a time dependent detuning $\Delta(t)$ and $\Delta=0$. 
The intracavity dynamics for the QE population $|c_e(t)|^2$ (solid lines) and the target cavity photon population $|c_t(t)|^2$ (dashed lines) are shown in Fig. 2 (b) for the three detuning plotted in Fig.2(a). As expected, even for a system with losses the above analysis holds i.e. the QE is completely decoupled in case of $\Delta=0$ and it decays as it would decay in an uncoupled cavity for large detuning $\Delta\gg\eta$. Due to the active modulation of the QE-cavity coupling via a time dependent detuning $\Delta(t)$ one can shape the emitted wavepacket in principle arbitrarily. Here, we show the shaping into an symmetric pulse, which can be absorbed with in principle unit fidelity by a similar system with time-inverted control cavity detuning. The output pulse is shown on Fig.2(c) together with the natural pulse shape without active control in case $\Delta\gg\eta$. A Gaussian fit of the shaped wavepacket reveals a nearly perfect time-symmetric wavepacket. The phase of the shaped wavepacket is shown in Fig.2(d) and it is constant in time (apart from the small deviations for long time due to numerical error caused by small amplitudes), since there is no frequency tuning involved in contrast to recent proposals using ultrafast electrical control of the QE energies \cite{Johne2011}. The fidelity $F$ of the absorption of this symmetric photon pulse by a similar system with time inverted operation can be obtained by calculating the overlap integral of the incident pulse with the time inverted pulse \cite{Johne2011}. This yields for the present case $F=0.997$ and can be further improved by optimization of the dynamic tuning.

The dynamic tuning of QE-cavity coupling requires that the coupled cavity modes follow the adiabatic eigenstates given by Eq.(2). Assuming a linear time dependence of the detuning $\Delta(t)=\beta t$, the problem can be described by a generalized Landau-Zener model \cite{carroll1986,Wang2006}.The resulting condition $\sqrt{\beta}\ll\eta$ needs to be satisfied to ensure adiabaticity. This sets an upper speed limit on the dynamic tuning. Including the QE-cavity coupling, the condition for shaping photon pulses can be written as $2{g^2}/\kappa  \ll \sqrt{\beta}  \ll \eta $, which is very well satisfied for the simulations shown in Fig.2. In case one opts for the switching of Rabi oscillations the more stringent condition reads $\kappa  < g \ll \sqrt{\beta}  \ll \eta $. On the other hand, detrimental off-resonant coupling of the QE to modes $\omega_{2,3}$ in case of $\Delta\approx0$ gives a lower speed limit governed by the ratio $
{g \mathord{\left/ {\vphantom {g \eta }} \right. \kern-\nulldelimiterspace} \eta }$, which should be small. However, there also the spontaneous decay of the QE in leaky modes $\gamma$ eventually may affect the performance. For both the lower and upper tuning speed limits, a large cavity coupling $\eta$ is desired and should be carefully engineered during the implementation. Note that while we focus here on the adiabatic tuning, diabatic effects may open up additional applications of the coupled-cavity system e.g. for unconventional beam splitters and for the generation of entanglement \cite{Quintana2013}.

The proposed scheme can be in principle implemented in any coupled-cavity QED system. Here we discuss two practical implementations, namely a semiconductor cavity consisting of quantum dots in a photonic crystal and and a microwave system using circuit QED.

In the solid state various implementations of coupled cavity systems have been realized such as microdiscs \cite{ryu2009,benyoucef2011}, nanowires \cite{bayer1998,galbiati2012}, ring resonators \cite{rabus2001} and photonic crystals \cite{notomi2007,vignolini2010,intonti2011,konoike2013}. Here we consider a coupled photonic crystal cavity-quantum dot system. The desired tuning can be implemented by modulation of two laser beams impinging on the control cavities injecting free carriers and thus providing the local refractive index change. The modulation speed is only limited by the free carrier lifetime in the material and can well exceed the SE time of a weakly coupled quantum dot \cite{jin2013}. By applying electric fields to reduce the free carrier lifetime down to a few picoseconds a modulation speed of about $100$ GHz can be reached.

To be more precise, we consider a system consisting of three in-line coupled L3 photonic crystal cavities. Three-dimensional finite element simulations are used to determine the local density of optical states (LDOS) $D$ experienced by an QE in the central target cavity depending on its wavelength for detunings of the control cavities as shown in Fig.3(a-c). The refractive index change can be translated into a frequency/wavelength change by using the expression $\Delta \omega /\omega  = -\Delta n/n$. For $\Delta=0$ the mode $\omega_1$ has no target cavity fraction and hence the dipole in the central cavity interacts only with the modes $\omega_{2,3}$ (Fig.3(a)). For intermediate detuning all three coupled modes have electric field distributions in the target cavity mode, each at its own frequency (Fig.3(b)). In case of very large detunings the central cavity mode is completely decoupled and only a single peak is visible in the LDOS spectrum (Fig.3(c)). The difference in the loss rates of modes $\omega_{2,3}$ arises from different diffractive out-of-plane losses \cite{atlasov2008}. The dependence of the LDOS at frequency $\omega_1$ on the detuning is shown in Fig.3(d). It is in good agreement with the simple coupled oscillator model ${D_t} = {|\alpha^{(1)} _t|^2}{D_0}$, where $D_0$ is the LDOS of the decoupled target cavity mode determined by the finite element calculations. Once again the LDOS experienced by the QE can be fully controlled by tuning the control cavities. The electric field profile of the mode $\omega_1$ is shown for three different detunings in Fig.3 (e). 
 
\begin{figure}
\begin{center}
\includegraphics[width=1.0\linewidth]{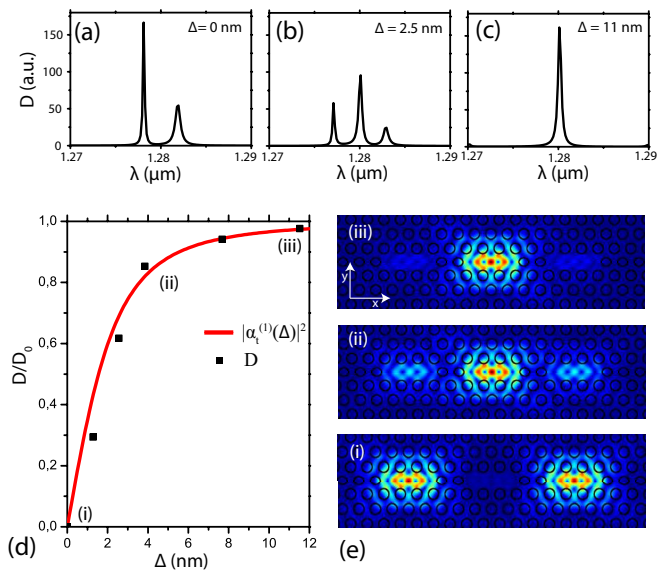}
\caption {\label{fig3} Spectra of the local density of optical states $D$ experienced by a dipole in the target cavity for a detuning (a) $\Delta=0$ nm (b) $\Delta=2.5$ nm (c) $\Delta=11$ nm.  (d) Normalized LDOS $D/D_0$ calculated by finite element simulations (squares) in comparison to the results from the coupled oscillator model $|\alpha^{(1)}_t|^2$ (red line). (e) Electric field distributions of the mode with frequency $\omega_1$ for three different detunings labelled (i-iii) indicated in (d).}
\end{center}
\end{figure}

An alternative implementation is possible in superconducting circuits \cite{devoret2013} due to the ultrahigh quality resonators as well as long coherence times of qubits. The development of tunable superconducting resonators \cite{wallquist2006,sandberg2008,palacios2008} has been the basis for tunable couplers with unprecedented level of control \cite{pierre2014}. This approach is also based on a coupled cavity system, where the frequency detuning of the of the resonators is controlled by a magnetic flux applied to a superconducting quantum interference device. Thus, circuit QED provides an ideal platform to realize the present proposal in the microwave regime.

Summarizing, the present proposal illustrates that a coupled-cavity system can be used to fully and deterministically control the QE-cavity coupling without inducing a spectral detuning between the QE and the cavity mode. This enables active shaping of single photon pulses in ultrafast experiments. Finally we propose an experimental design, which can be used to implement the tuning scheme. The present proposal provides a powerful technique to control the SE of QEs and paves the way towards a fully controllable single photon source. Furthermore, the results show the potential of the coupled cavity approach for actively controlling the light-matter interaction in the solid state enabling more advanced application in quantum information processing.  

The authors acknowledge enlightening discussions on coupled cavities with Massimo Gurioli (Univ. Firenze) and assistance from Leonardo Midolo in the finite element simulations. This work is part of the research programme of the Foundation for Fundamental Research on Matter (FOM), which is financially supported by the Netherlands Organization for Scientific Research (NWO), and is also supported by the Dutch Technology Foundation STW, applied science division of NWO, the Technology Program of the Ministry of Economic Affairs under Project No. 10380.

\end{document}